\documentclass[conference]{IEEEtran}
\usepackage{amsmath,amssymb,graphicx,relsize}
\usepackage{hyperref}
\usepackage{graphicx}
\usepackage{booktabs} 
\usepackage{svg}
\usepackage{hyperref}
\usepackage{cleveref}

\title{Dysarthria Normalization via Local Lie Group Transformations for Robust ASR}

\author{
    \IEEEauthorblockN{Mikhail Osipov}
    \IEEEauthorblockA{Independent Researcher, Milan, Italy \\ \texttt{osipov.ma@phystech.edu}}
}

\begin{document}


\maketitle

\begin{abstract}
We present a geometry-driven method for normalizing dysarthric speech by modeling time, frequency, and amplitude distortions as smooth, local Lie group transformations of spectrograms. Scalar fields generate these deformations via exponential maps, and a neural network is trained—using only synthetically warped healthy speech—to infer the fields and apply an approximate inverse at test time. We introduce a spontaneous-symmetry-breaking (SSB) potential that encourages the model to discover non-trivial field configurations, preventing collapse and enabling stable training under strong synthetic distortions. On real pathological speech, the system delivers consistent gains: up to 17 percentage-point WER reduction on challenging TORGO utterances and a 16\% drop in WER variance, with no degradation on clean CommonVoice data. Character and phoneme error rates improve in parallel, confirming linguistic relevance. Random but equally smooth Lie-type warps improve less, showing that input-conditioned field prediction matters. Our results demonstrate that geometrically structured warping provides consistent, zero-shot robustness gains for dysarthric ASR.
\end{abstract}

\section{Introduction}

Automatic speech recognition (ASR) systems have reached impressive levels of accuracy on typical, fluent speech. However, their performance significantly degrades when applied to dysarthric speech, a motor speech disorder caused by neurological impairments that affects articulatory precision, timing, and prosody. Dysarthric speech often includes irregular pacing, slurred phonemes, reduced volume, and unstable frequency patterns---all of which pose difficulties for conventional ASR models trained on healthy speech data \cite{kim2008uasp, rudzicz2012torgo}.

Existing approaches to improving dysarthric ASR fall into several categories: data augmentation \cite{park2019specaugment}, speech enhancement, and voice conversion. While data augmentation attempts to make ASR more robust to acoustic variation, it lacks the specificity to model the structured nature of dysarthric distortions. Speech enhancement and voice conversion methods often operate as black boxes, providing limited interpretability and typically relying on large amounts of paired data, which is difficult to collect for clinical populations.

We propose a new perspective: to treat dysarthric speech as a \textit{transformed version} of typical speech in the time--frequency domain. Motivated by mostly continuous and local nature of dysarthric distortions, we model these transformations as elements of a Lie group acting on spectrograms. Each transformation---such as local time stretching or frequency scaling---is generated by a corresponding Lie algebra operator and parameterized by a smooth scalar field. This structured approach allows us to simulate dysarthric speech in a principled way and learn to invert such transformations from data \cite{ardila2020common}.

Empirical results show that even when applied to clean speech, the predicted transformations do not degrade recognition accuracy---suggesting the model learns input-conditioned, linguistically consistent fields. Furthermore, although predicted fields have similar magnitude across normal and dysarthric inputs, their effect is corrective primarily on disordered samples. This supports the interpretation that learned transformations serve as a robust front-end, not merely a blind inversion mechanism.

\textbf{Our contributions are as follows:}
\begin{enumerate}
\item We introduce a novel Lie-group framework for modeling dysarthric distortions as local spectrogram transformations, including time stretch, frequency stretch, 2D warp, and amplitude modulation.
\item We propose a training strategy that includes a spontaneous-symmetry-breaking (SSB) potential and a masked local loss, encouraging the emergence of non-trivial, spatially selective transformation fields.
\item We demonstrate that the model improves ASR performance on real dysarthric speech (TORGO and UA-Speech), particularly on high-WER cases, while preserving accuracy on clean speech.
\item We show that phoneme-level recognition improves alongside WER and CER, and that the variance of WER across test samples is significantly reduced.
\item We compare to both simpler ablations and randomly applied fields, showing that structured, learned warps yield significantly greater improvements.
\item We confirm these gains using multiple ASR backends (RNNT, Whisper), demonstrating that our front-end generalizes across architectures.
\end{enumerate}

This work opens a path toward interpretable, transformation-aware front-ends for dysarthric speech recognition and suggests new avenues for bridging speech disorder modeling with geometric machine learning. Furthermore, our formulation conceptually connects to recent developments in several emerging domains.

First, the forward application and inverse removal of spectro-temporal distortions resembles a diffusion process in \textit{diffusion models}, where a clean spectrogram is progressively perturbed and then denoised. Our transformation operators can be viewed as structured alternatives to Gaussian noise, with potential for integration into score-based generative modeling \cite{huang2022diffusion}.

Also, the transformations we apply resemble continuous deformations of a physical system. Our scalar fields could be regularized or governed by constraints inspired by speech production physics, such as bounded articulator velocity or energy conservation, similar to what is happening in \textit{physics-aware neural networks} \cite{grathwohl2019ffjord, arjovsky2019neural}.

Finally, the scalar fields parameterizing local transformations over the spectrogram can be interpreted as realizations of \textit{random fields}. This opens connections to Gaussian processes and spatial statistics, providing a formal basis for sampling and modeling realistic dysarthric distortion patterns \cite{younes2019shape, bronstein2021geometric}.

These connections offer theoretical insight and practical opportunities to integrate structured priors, generative mechanisms, and physical constraints into speech processing for disordered speech.

\section{Related Work}

Dysarthric speech recognition has long been a difficult problem for ASR systems, largely due to the variability and irregularity introduced by motor impairments. Datasets such as UA-Speech \cite{kim2008uasp} and TORGO \cite{rudzicz2012torgo} have served as benchmarks for evaluating ASR models under dysarthric conditions. However, even state-of-the-art systems exhibit significantly degraded performance on these datasets, especially for speakers with severe articulation impairments. A common observation is that standard ASR models fail to generalize due to mismatches in timing, spectral features, and phoneme realization.

To mitigate these issues, several strategies have emerged. Recent advances in voice conversion (VC) have leveraged adversarial and cycle-consistent learning to enable non-parallel speech domain mapping. Models such as CycleGAN-VC \cite{kaneko2018cyclegan} and its successors rely on generative losses without explicit correspondence to the underlying physiological or articulatory mechanisms of speech. This trend continues in modern dysarthric speech processing, where adversarial VC models have been applied to convert disordered speech to a canonical form using cycle-consistency, GAN-based realism objectives, or autoencoding pipelines \cite{li2024unidiffusion, zhang2024voiceidvc, tan2020improving}. While these methods demonstrate strong empirical performance, they often operate as black boxes—lacking interpretability, invertibility, and grounding in known patterns of speech distortion.

In parallel, a body of work has explored spectrogram-space manipulations as a means to achieve robustness. SpecAugment \cite{tan2020improving} introduces frequency and time masking and warping directly in the spectrogram domain, improving performance on many tasks. Extensions of these techniques have been used for dysarthric speech as well, introducing modifications that mimic speech degradation. However, such augmentations are typically heuristic and do not reflect an underlying generative model of the disorder. Other studies have investigated frequency warping or temporal alignment strategies \cite{10095239,10248099}, but often in an isolated fashion, lacking a cohesive model to represent the interplay of multiple distortions.

Recently, researchers have begun to explore more structured geometric representations of speech transformations. Vocal tract length normalization (VTLN) has been formulated using Lie group theory, where frequency warping corresponds to an all-pass transformation within a continuous group structure \cite{10248099}. Similarly, general time-warping operators have been modeled as diffeomorphic transformations or as Lie group elements with differentiable representations \cite{younes2019shape, bronstein2021geometric}. These works lay the mathematical foundation for treating speech variation not as noise, but as a structured transformation that can be learned and inverted.

Our work draws inspiration from some of these advances, but goes beyond them by making an assumption that \textit{dysarthric distortions can be locally modelled with a Lie group of spectrogram transformations}. We integrate multiple types of transformations—time stretch, frequency scaling, amplitude modulation, and 2D warping—into a unified Lie group framework and associate each transformation with a local scalar \textit{random field} over the spectrogram and learn to estimate these fields from data. This enables not only reconstruction and normalization but also visualization and interpretability of the distortions, bridging the gap between theory-driven modeling and practical ASR improvements.

\section{Method}

We introduce a geometric framework that models dysarthric speech distortions as local, structured transformations of a clean spectrogram. These transformations—such as time stretching, frequency scaling, amplitude modulation, and 2D spatial warping—are treated as elements of a continuous Lie group acting on the time--frequency domain. Each transformation is generated by a corresponding Lie algebra operator and is parameterized by a scalar field defined over the spectrogram grid. To learn and invert these transformations, we construct synthetic training pairs by applying known distortions to clean speech and train a neural network to estimate the underlying transformation fields. The predicted fields are then used to compute an approximate inverse transformation, effectively ``undoing'' the distortions and recovering a normalized spectrogram suitable for robust ASR.

We operate in the time--frequency domain using the short-time Fourier transform (STFT), denoted as:
\[
S(f,t) = M(f,t) \, e^{-i \varphi(f,t)},
\]
where $S(f,t)$ is the complex-valued STFT of a speech signal, $M(f,t)$ is its magnitude, and $\varphi(f,t)$ is the phase. Dysarthric distortions are modeled as transformations acting directly on $S(f,t)$.

\subsection{Time--Frequency Warping with Complex Scaling}

We define a general parametric transformation $T$ applied to a spectrogram as:
\[
\widetilde{S}(f,t) = \rho \, e^{-i \beta} \, S\big(\omega, \tau\big),
\]
where:
\begin{itemize}
    \item $\tau = \tau(f,t) \in \mathbb{R}$ is a smooth, invertible mapping of the time axis (local time warping),
    \item $\omega = \omega(f,t) \in \mathbb{R}$ is a smooth, invertible mapping of the frequency axis (local frequency warping),
    \item $\rho = \rho(f,t) \in \mathbb{R}_{+}$ is a local magnitude scaling factor,
    \item $\beta = \beta(f,t) \in \mathbb{R}$ is a local phase offset.
\end{itemize}

This formulation encapsulates a broad class of speech distortions, especially those observed in dysarthria \cite{kim2008uasp, rudzicz2012torgo, 10248099}. It is invertible under mild regularity conditions, assuming smooth scalar fields and non-degenerate coordinate mappings \cite{younes2019shape}.

\begin{itemize}
    \item $\tau$ and $\omega$ are diffeomorphisms,
    \item $\rho(f,t) \neq 0$ everywhere,
    \item $\beta(f,t)$ is real-valued and smooth.
\end{itemize}

This transformation model aims to represent:
\begin{itemize}
    \item \textbf{Time warping} $\tau(f,t)$: irregular pacing, segment prolongation or compression,
    \item \textbf{Frequency warping} $\omega(f,t)$: formant shifts, slurring or smearing of spectral content,
    \item \textbf{Amplitude scaling} $\rho(f,t)$: local weakening, variable vocal effort,
    \item \textbf{Phase shifts} $\beta(f,t)$: onset misalignments or pitch/voicing deviations.
\end{itemize}

\subsection{Infinite-Dimensional Group Structure}

The transformation classes described above form \textit{infinite-dimensional Lie groups}, as each point in the time-frequency space $(f, t)$ is associated with its own local parameters. For instance, time warping alone—defined via smooth, invertible maps $t \mapsto \tau(t)$—constitutes an infinite-dimensional group of diffeomorphisms. Similarly, pointwise complex scaling transformations of the form $S(f,t) \mapsto \rho(f,t)e^{-i \beta(f,t)} \, S(f,t)$, with $\beta \in \mathbb{R}, \rho\in \mathbb{R}_{+}$, form a commutative infinite-dimensional group under pointwise multiplication.

Each transformation admits an associated Lie algebra, consisting of infinitesimal generators \cite{youn2021group}. While this formulation assumes smooth invertibility, it excludes operations such as hard clipping, blur, or discontinuous phase manipulation, which fall outside the Lie group structure. Such non-invertible effects may arise in dysarthric speech (e.g., dropped phonemes or clipped bursts) \cite{kim2008uasp, rudzicz2012torgo}, and can be added to the machine learning pipeline, but they are not modeled directly in this framework.

\subsection{Properties of Transformations}

When modeling dysarthric speech as transformations of the complex-valued spectrogram, we aim for transformations that are not only mathematically well-defined, but also \textit{physically plausible}. In particular, the transformation fields should reflect realistic articulatory dynamics—smooth in time and frequency, with gradual changes that mirror how human vocal tract motion distorts speech \cite{younes2019shape, bronstein2021geometric}. For example, time warps should not introduce discontinuities, and amplitude or phase shifts should reflect feasible modulations of speech effort and voicing \cite{kim2008uasp, rudzicz2012torgo}. These constraints guide both the design of synthetic transformations for training and the regularization strategies used during learning.

To ensure the transformations remain realistic and interpretable, we introduce constraints reflecting typical patterns in dysarthric speech. 

First, time warping should preserve the natural forward progression of time. This motivates imposing a monotonicity constraint: $\tau'(t) > 0$.

Second, frequency warping should preserve the continuity of spectral content. Real vocal tract deformations shift or smear formants smoothly \cite{fant1970acoustic, kim2008uasp}. Thus, the mapping \( \omega(f,t) \) should avoid sharp discontinuities or folding, ensuring gradual spectral changes \cite{10248099, youn2021group}.

Third, distortions in dysarthric speech are often localized—certain phonemes or time–frequency regions may be more affected than others \cite{rudzicz2012torgo, kim2008uasp}. We incorporate this by encouraging the amplitude and phase fields \( \rho(f,t) \) and \( \beta(f,t) \) to stay close to identity values (i.e., \( \rho \approx 1 \), \( \beta \approx 0 \)) across most of the spectrogram, allowing larger deviations only in localized regions. This reflects the fact that dysarthria typically alters speech selectively, rather than uniformly across all frequencies and times \cite{ball2004introduction}.

Another important modeling question is whether transformations should preserve the total energy of the spectrogram, i.e.,
\[
\int |S(f,t)|^2 \, df \, dt = \int |\widetilde{S}(f,t)|^2 \, df \, dt.
\]
While strict energy preservation is desirable in settings aiming to model lossless changes in vocal tract shaping or to maintain physical consistency \cite{fant1970acoustic}, real dysarthric speech often exhibits genuine energy loss. This is especially true in hypokinetic dysarthria, where speakers produce reduced loudness and weakened consonant bursts \cite{ball2004introduction, darley1969dysarthria}. In such cases, enforcing global energy conservation would be overly restrictive. Instead, we allow amplitude scaling transformations that can locally reduce energy, reflecting realistic vocal effort variations. Our model tracks and optionally penalizes excessive energy change but does not enforce hard conservation.

\subsection{Lie Algebra Generators}

We define a family of infinitesimal generators that describe local spectrogram transformations in the STFT domain. These generators form the basis of a Lie algebra, whose exponentiation yields smooth, invertible time--frequency deformations commonly observed in dysarthric speech.

Specifically, we introduce:

\begin{itemize}
    \item $v(t)$: global time warp field,
    \item $w(f)$: global frequency warp field,
    \item $u_t(f,t)$: localized 2D time-warp component,
    \item $u_f(f,t)$: localized 2D frequency-warp component,
    \item $\alpha(f,t)$: amplitude modulation field,
    \item $\beta(f,t)$: phase modulation field.
\end{itemize}

A general infinitesimal generator acting on $S(f,t)$ is then given by:
\[
X = \big(v + u_t\big) \frac{\partial}{\partial t} + \big(w + u_f\big) \frac{\partial}{\partial f}
+ \alpha + i \beta.
\]

Exponentiating $X$ produces a finite transformation:
\[
\widetilde S = \exp(\varepsilon X)[S], \quad \text{with} \quad \widetilde S \approx S + \varepsilon X[S] + \mathcal{O}(\varepsilon^2).
\]

This Lie algebra enables the model to represent both global speech dynamics (e.g., uniform slowing, pitch shifts) and local distortions (e.g., phoneme-specific smearing or intensity variation), which are common in dysarthric speech. We discuss its mathematical properties in Appendix~\ref{app:algebra_properties}.

\textit{Note:} In practice, we observe that the learned 2D warp fields $(u_t, u_f)$ often subsume the effects of separate time and frequency scalar warps $v, w$ when used together in training. Ablation studies suggest that 2D warping may “cannibalize” or dominate the simpler one-dimensional warps—or vice versa—depending on the training schedule and hyperparameter configuration. Nonetheless, we retain all generators in the formulation for theoretical completeness and for potential future extensions requiring disentangled or controllable transformations.

\subsection{Discretization}

In this work, we restrict our transformations to act only on the magnitude of the spectrogram, ignoring the phase component. Mathematically, this corresponds to setting the local phase modulation field \( \beta(f,t) = 0 \), which defines a closed Lie subgroup of the full complex-valued transformation group:
\[
S(f,t) \longrightarrow |S(f,t)| \in \mathbb{R}_{+}
\]
This choice preserves interpretability and stability, and avoids the challenges of phase modeling in the short-time Fourier transform (STFT) domain. It also aligns with common practice in ASR systems, which typically operate on magnitude or log-mel features. By focusing on the real-valued subgroup, we ensure compatibility with existing pipelines while still capturing rich time–frequency distortions through smooth coordinate warps and amplitude modulations.

We discretize the spectrogram into a grid of shape \( F \times T \), where:
\begin{itemize}
    \item \( F \) is the number of frequency bins (e.g., 80 mel bands),
    \item \( T \) is the number of time frames (e.g., 512 for $\approx 5$ seconds of audio).
\end{itemize}

Then, we define the following fields:
\begin{itemize}
    \item \( \phi_{\text{time}}(t) \in \mathbb{R}^{T} \): 1D global time warp field, 
    \item \( \phi_{\text{freq}}(f) \in \mathbb{R}^{F} \): 1D global frequency warp field,
    \item \( \phi_{u_t}(f,t), \phi_{u_f}(f,t) \in \mathbb{R}^{F \times T} \): 2D local warping fields,
    \item \( \phi_{\text{amp}}(f,t) \in \mathbb{R}^{F \times T} \): amplitude modulation field.
\end{itemize}

After broad-casting $\phi_{\text{time}}$ along $f$ axis and broad-casting $\phi_{\text{freq}}$ along $t$ axis, each scalar field becomes represented as a real-valued matrix of shape \( F \times T \), aligned with the spectrogram grid.\footnote{One-dimensional fields $\phi_{\text{time}}$ and $\phi_{\text{freq}}$ are represented by 2D matrix, which is not efficient for ML task; we use this approach for simplicity. Importantly, we can generate separately time stretching, frequency stretching and combined 2D warps during forward pass.}

These fields are the primary objects of prediction in our model. During training, we sample synthetic fields with known parameters, apply the corresponding transformations to clean spectrograms, and train a neural network to estimate the fields from the transformed data. During inference, the predicted fields are used to compute an approximate inverse transformation.

The total dimensionality of the generated scalar fields is given by:
\[
\dim(\phi) = T + F + 3 \cdot (F \times T),
\]
accounting for one global time warp field, one global frequency warp field, and three dense 2D fields (local time warp, local frequency warp, and amplitude modulation), all defined over a spectrogram grid of shape \( F \times T \).

\textit{Note:} One can reduce the class of fields and/or make use of a sparse control grid to represent transformation parameters more compactly. This idea is outlined in Appendix~\ref{app:sparse_fields}. In our current experiments, we use dense, smooth local fields during training for simplicity, but sparse interpolation schemes could significantly reduce parameter count and improve training efficiency in future implementations. Also, for practical implementation, we apply transformations using a first-order approximation of the exponential map. However, the underlying Lie group structure naturally supports large distortions via higher-order or iterative exponential flows, as discussed in Appendix~\ref{app:nonlinear_expansion}. While not used during training in this work, these expansions offer a principled way to model more severe dysarthric distortions without increasing field dimensionality.

\subsection{Synthetic Transformation Fields}

To train the model in a supervised setting, we generate synthetic transformation fields and apply them to clean spectrograms of typical speech. These fields are sampled as smooth, localized sinusoidal patterns designed to mimic real dysarthric distortions in time, frequency, and amplitude.

We generate 1D global warp fields using localized sinusoidal blobs masked by soft Gaussians. Similarly, 2D local fields are constructed by superimposing a small number of spatially confined sinusoidal waves in the time--frequency plane. This yields realistic, structured distortions with controllable smoothness and amplitude. Full details of the field generators are provided in Appendix~\ref{app:field_generation}.

\begin{figure}[ht!]
\centering
\includegraphics[width=80mm]{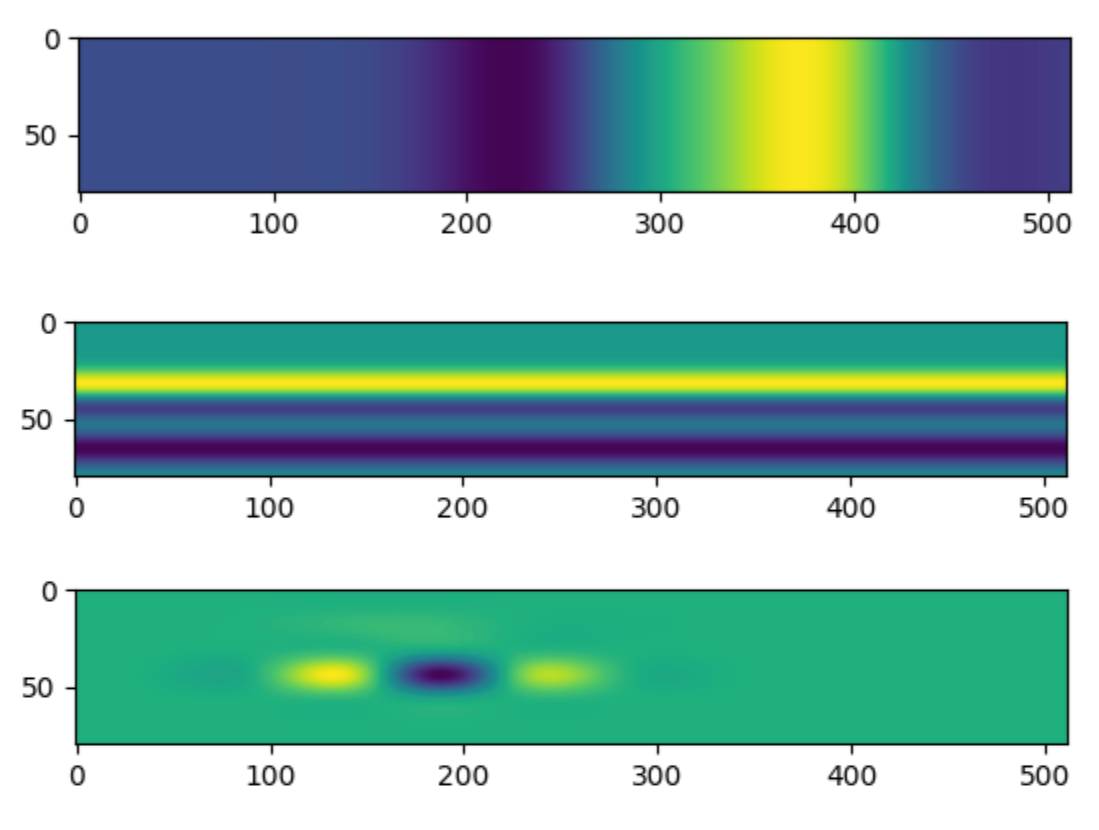}
\caption{Examples of generated fields: $\phi_{\text{time}}$, $\phi_{\text{freq}}$ (before masking), $\phi_{\text{amp}}$ (masked)}
\end{figure}

\subsection{Applying Spectrogram Transformations for Training}

To generate training data, we apply randomly sampled smooth transformation fields to clean spectrograms from the CommonVoice \cite{ardila2020common} dataset. These fields are drawn from a parametric Lie algebra (see Appendix~\ref{app:field_generation}) and represent synthetic dysarthria-like distortions.

Each transformation is applied using a deformation field generated at runtime and passed to a differentiable warping function (bilinear interpolation over a coordinate grid). The magnitude of each transformation is controlled via a dictionary of values that specifies maximum perturbation levels for each generator field. The relative scale of values that correspond to different transformations was defined experimentally to result in comparable loss values.

To encourage stable training and gradual adaptation, we implement a simple curriculum learning schedule. Early in training, only identity-like ($\varepsilon \ll 1$) distortions are used. Over time, the values of $\varepsilon$ are increased, allowing the model to see progressively more challenging examples. While the framework supports composing multiple transformations per sample, in our current setup we apply only one transformation at a time. This simplification was made for code clarity and training stability, though multi-transform compositions are expected to further improve model robustness.

\subsection{Approximate Inverse Transformation}

After predicting the transformation fields from a distorted spectrogram, we apply an approximate inverse operation to recover a cleaned version of the original signal. Each predicted field is first rescaled from a normalized range \([-1, 1]\) to its true deformation range using a predefined dictionary of values\footnote{To apply the inverse transformation, we use the same set of scaling factors $\varepsilon$ that was used for the $\phi$ fields generation. That means that the model learns the spatial distribution of the fields, but we guide it providing the overall scale.}. The inverse operation consists of reversing the time and frequency warps, inverting the 2D deformation flow, and undoing amplitude modulation via safe division. This process yields a reconstructed spectrogram that approximates the original undistorted input. Full implementation details and tensor shapes are provided in Appendix~\ref{app:inverse_transform}.

\subsection{Interpretability and Non-Uniqueness of Inversion}

For any distorted spectrogram, there exists an infinite number of possible Lie generator fields that could have produced it—owing to the non-commutative, compositional nature of the transformation group \cite{younes2019shape, bronstein2021geometric}. Our model does not aim to find the “true” inverse in a literal sense. Instead, it learns to infer a consistent and plausible set of local fields that, when inverted, approximate a cleaner spectrogram and improve ASR performance. This perspective aligns with general principles in inverse problems and generative modeling, where the goal is not unique recovery, but structured, useful inversion \cite{grathwohl2019ffjord, arjovsky2019neural}. \footnote{This is conceptually similar to diffusion models and score-based methods, where a single noisy input may correspond to many latent states, and the objective is to recover one that is both consistent and beneficial for downstream tasks \cite{huang2022diffusion}}.

During training, the synthetic transformations are applied randomly and independently of phonetic structure. Thus, the model cannot rely on any external alignment or prior. The only viable strategy is to learn to detect undistorted (or typical) spectral patterns and propose local inverse transformations where deviations occur. Over time, the model is supposed to internalize structural regularities of speech—such as formant contours and harmonic stacks—and learn to “explain away” distortions by generating suitable Lie fields. 

This process echoes the principles of unsupervised speech learning and denoising representation models, where the system must infer structure purely from consistency and reconstruction feedback \cite{oord2018representation, baevski2020wav2vec}. The learned inverse is therefore one of many valid interpretations — or, in other terms, one of the realizations of the random fields \( \hat\phi(f,t) \) — sufficient as long as it yields improved reconstructions and better downstream recognition. This idea parallels probabilistic inversion and self-consistency frameworks widely used in vision and graphics, where models learn to undo distortions or reconstruct hidden structure without requiring exact supervision \cite{tulsiani2017multi, zhou2016view, rezende2014stochastic}.

\section{Experiments}

\subsection{Datasets}

For training, we use the English portion of the Common Voice dataset (v17.0) \cite{ardila2020common} as a source of clean, healthy speech. Synthetic spectrogram distortions are generated from this data using randomly sampled Lie transformation fields, without any reliance on phonetic labels or alignment. For evaluation, we use the “dysarthric” subset of the TORGO dataset \cite{rudzicz2012torgo}, which contains real dysarthric speech from speakers with varying severity levels and etiologies, and a subset of UA-Speech\cite{kim2008uasp} dataset with "medium" dysarthria severity labels. Importantly, our model is never exposed to TORGO, UA-Speech or any pathological speech data during training. This zero-shot setup highlights the model’s ability to generalize from synthetic to real-world distortions. As the training procedure is language-agnostic and depends only on spectrogram structure, the approach can be trivially extended to other languages with available ASR systems.

\subsection{Neural Architecture: U-Net with ResNeXt Backbone}

To predict the transformation fields from spectrograms, we use a U-Net architecture with a ResNeXt-50 encoder, implemented via the \texttt{smp.Unet} framework. Our choice balances robustness, efficiency, and compatibility with image-like spectrogram data.

We use a single-channel input (magnitude spectrogram) and predict five scalar fields (local time and frequency warps, amplitude, and global warp terms). The pretrained ImageNet weights provide strong initial features, while the U-Net architecture ensures both global context and fine-grained localization—crucial for recovering smooth, localized deformation fields without losing key spectral detail. Detailed information on the pre-trained model is available in Appendix~\ref{app:nnet}.

The use of encoder-decoder structures like U-Net has proven effective for dense prediction tasks across domains, including speech-related applications \cite{ronneberger2015u}.

\subsection{Loss Function}

Our training objective combines multiple complementary terms to guide the model toward accurate field prediction, faithful spectrogram reconstruction, and physically plausible deformation patterns. The initial loss function used in early experiments consisted of four components: (1) a mean squared error (MSE) loss between predicted and ground-truth fields, (2) a cosine similarity loss to encourage directional alignment of vector-valued fields, (3) an MSE reconstruction loss comparing the inverse-transformed and clean spectrograms, and (4) a spatial smoothness penalty on the predicted fields to ensure regularity.

In the improved formulation used for our final model, we introduced two additional components. First, we added a \textit{spontaneous-symmetry-breaking} (SSB) potential in the form of a “hat” loss, encouraging the norm of each predicted field to approach a characteristic value  $\hat{\phi}$ that adapts to the ground-truth field norm—thereby promoting non-trivial field configurations and preventing collapse. Second, we incorporated an L1 sparsity loss selectively applied to a subset of fields (time stretch and 2D warp), reflecting empirical observations that these fields tend to exhibit spurious activation on silent or irrelevant regions. The loss function $\mathcal{L}[\phi_{\text{pred}},\phi_{\text{true}},S_{\text{true}}] $ becomes similar to a field lagrangian:
\[
\mathcal{L} = \sum\limits_{\lambda \in \Lambda} \lambda\cdot\mathcal L_{\lambda}, ~~ \Lambda = \big\{\text{spec},\text{cosine},\text{kinetic},\text{ssb},\text{sparse}\big\}.
\]
The loss terms are given by:
\begin{multline}\label{eq:1}
\mathcal{L}_{\text{spec}} = \sum\limits_{f,t \in \Omega} \left(S_{\text{pred}} - S_{\text{true}}\right)^2 = \sum\limits_{f,t \in \Omega} \left( ( \exp(\varepsilon X)-1) [S_{\text{true}}] \right)^2,\\
\\
\mathcal{L}_{\text{ssb}} = \sum\limits_{f,t \in \Theta} \left( \left\| \phi_{\text{pred}} \right\| - \overline{\phi_{\text{true}}} \right)^2 + \sum\limits_{f,t \in \Omega \setminus \Theta}\left\| \phi_{\text{pred}} \right \|^2,\\
\\
\mathcal{L}_{\text{cosine}} = \sum\limits_{f,t \in \Omega} \frac{\phi_{\text{pred}}\cdot \phi_{\text{true}}}{\left\| \phi_{\text{pred}} \right\|^2 \left\| \phi_{\text{true}} \right\|^2}\\
\mathcal{L}_{\text{kinetic}} = \sum\limits_{f,t \in \Omega}  \left\| \frac{\partial \phi}{\partial f} \right\|^2 + \left\| \frac{\partial \phi}{\partial t} \right\|^2,\\
\mathcal{L}_{\text{sparse}}  = \sum\limits_{f,t \in \Omega} \left\| \phi_{\text{pred}}\right \|_1.\\
\end{multline}

By $\Omega$ we denote the discretized $F \times T$ space, with the region of non-zero true field values $\Theta = \{ f,t: \phi_{\text{true}}(f,t) > 0 \}$. Each term in the final loss is weighted by a tunable coefficient $\lambda$. 

Additionally, the framework allows for the inclusion of more perceptually grounded losses, such as Mel Cepstral Distortion (MCD) for audio fidelity \cite{kubichek1993mel} or ASR-based perceptual loss computed on model transcriptions \cite{huang2022diffusion, takaki2021end}. These were not used in our current experiments but could further align reconstruction quality with downstream recognition performance.

\subsection{Training}

The initial version of the model (v1) was trained for 10 epochs on a 10{,}000-sample subset of the English CommonVoice dataset (v17.0), using distributed data parallelism (DDP) across 4$\times$ NVIDIA GeForce RTX 4090 GPUs. We employed an exponential learning rate scheduler alongside a simple curriculum schedule based on the $\varepsilon$ parameter, which controls the strength of synthetic spectrogram distortions. This schedule consisted of a linear warm-up phase, a plateau of moderate difficulty, and a final linear increase to stronger distortions (see Fig.~\ref{fig:train_loss_v1} in Appendix~\ref{app:training_v1}).

While the training was stable, the v1 model frequently collapsed into a trivial solution with near-zero predicted transformation fields, especially under moderate-to-strong distortion regimes. Despite some improvement in ASR performance, this behavior suggested the model failed to fully leverage its capacity to discover useful, nontrivial transformations.

\begin{figure}[ht!]
\centering
\includegraphics[width=80mm]{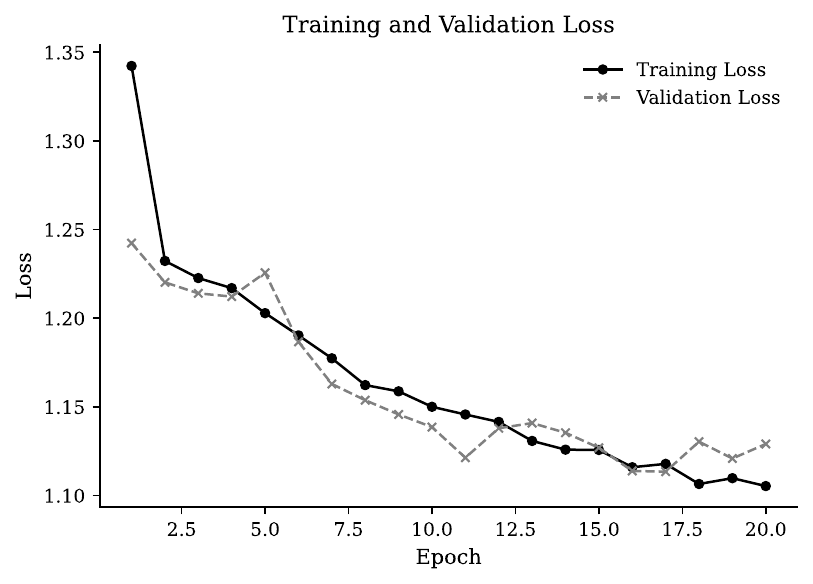}
\caption{Training and validation loss dynamics (10000 samples from CommonVoice dataset. Model version v.2, batch size = 16, learning rate starts from $3 e^{-5}$). The $\varepsilon$ parameter grows linearly on warmup stage, plateaus and then grows linearly until the end of training}
\label{fig:train2_epochs}
\end{figure}

\begin{figure}[ht!]
\centering
\includegraphics[width=80mm]{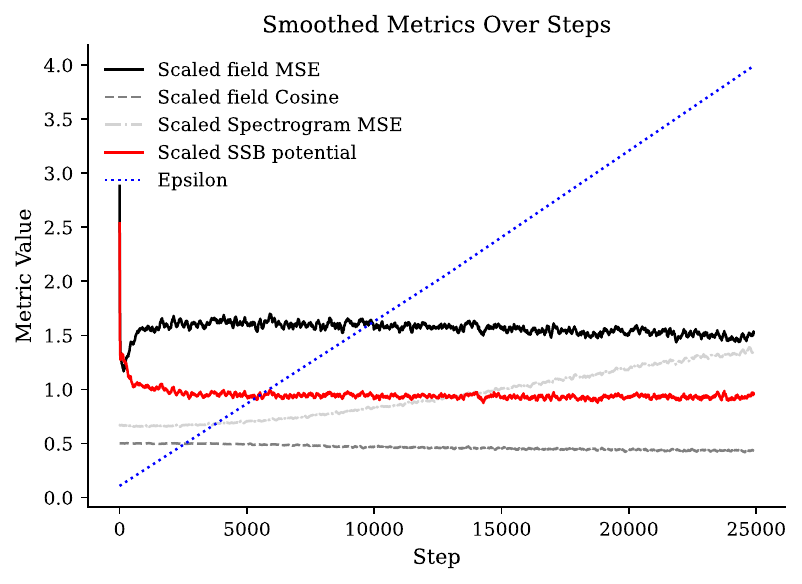}
\caption{Scaled loss function terms and $\varepsilon$ dynamics across training steps. Model version v.2}
\label{fig:train2_steps}
\end{figure}

To address this, we introduced a spontaneous symmetry breaking (SSB) potential in the loss function (\ref{eq:1}), encouraging the model to escape the trivial vacuum and converge toward richer, structured field configurations. This change was accompanied by a masked loss formulation, where the field alignment terms were applied only in regions of significant deformation, and a sparsity penalty selectively applied to time and 2D warping components. The updated model (v2) was trained for 20 epochs (Fig. \ref{fig:train2_epochs}) with a simpler curriculum:  $\varepsilon$ was scaled linearly from 0.1 to 4.0 with no intermediate plateau (Fig. \ref{fig:train2_steps}). This resulted in more stable optimization, higher magnitude and spatial structure in the learned fields, and overall better ASR performance. 

\subsection{ASR-based Evaluation}

To evaluate downstream performance on the ASR task, we used three test sets of 5{,}000 samples each: (1) dysarthric speech from the TORGO dataset, (2) speech from the UASpeech dataset with medium dysarthria severity labels, and (3) a control set of fluent speech from the English CommonVoice dataset (v17.0). Importantly, none of these samples were seen by the model during training or validation.

We used the large English Conformer-Transducer model from NVIDIA NeMo\cite{kuchaiev2019nemo} and CrisperWhisper version of OpenAI Whisper (\cite{radford2022robustspeechrecognitionlargescale,wagner2024crisperwhisperaccuratetimestampsverbatim}) as our ASR backends. We selected CrisperWhisper as an additional ASR backend due to its training objective: unlike standard Whisper models, it is optimized for verbatim transcription, reducing paraphrasing and hallucinations. This makes it especially suitable for dysarthric datasets, which consist of short, clearly segmented utterances where literal transcription fidelity is preferred. During inference, we configured decoding to be deterministic and conservative to further reduce hallucinated content and stabilize evaluation.  

For each input spectrogram, we applied our model to predict the transformation fields and then reconstructed a corrected spectrogram via the inverse Lie transformation. Both the original and reconstructed spectrograms were independently passed to the ASR model, and we recorded word error rate (WER), phoneme error rate (PER) and character error rate (CER) for each version.

Evaluation results are summarized in Table~\ref{tab:wer2}, which reports WER and PER (in percentage) for two input types: the unmodified spectrogram (marked as "identity" transformation), and the spectrogram corrected using model version 2 (v2). \footnote{
\textit{Evaluation Note.} To accommodate different ASR backends and dataset characteristics, we used varying transformation strength values (denoted as $\varepsilon$) during inference. Specifically:
\begin{itemize}
    \item RNNT–Torgo: $\varepsilon$ = 0.6
    \item RNNT–UASpeech: $\varepsilon$ = 2.0
    \item RNNT–CommonVoice: $\varepsilon$ = 0.5
    \item Whisper (CrisperWhisper)–All: $\varepsilon$ = 3.0
\end{itemize}
These values were chosen empirically to balance transformation strength with ASR performance across conditions.}

To analyze model performance in more challenging conditions, we also report in Table~\ref{tab:wer50} WER and PER on a “problematic” subset of each dataset, defined as follows:
\begin{itemize}
    \item For WER: samples where the ASR model achieves WER~$>$~50\% on the original spectrogram.
    \item For PER: samples where the ASR model achieves PER~$>$~30\% on the original spectrogram.
\end{itemize}

For each improvement observed, we conducted a paired Wilcoxon signed-rank test to assess statistical significance. All reported improvements yielded $p$-values $\ll 0.05$, confirming that the gains were statistically significant.

Evaluation metrics obtained with the model version 1 (v1) are available in Appendix~\ref{app:results_v1}.

\begin{table}[h]
\centering
\caption{WER, WER variance and PER across different datasets and spectrogram transformations}
\label{tab:wer2}
\vspace{0.5em}
\begin{tabular}{l|cc|cc}
\toprule
\textbf{} & \multicolumn{2}{c|} {\textbf{RNNT-Conformer}} & \multicolumn{2}{c} {\textbf{CrisperWhisper}}  \\
& \textit{Identity} & \textit{Model v.2} & \textit{Identity} & \textit{Model v.2} \\         
\midrule
\textbf{WER, \%} \\
TORGO$_{5000}$          & 77.3     & 72.3      & 102.8          & 92.9 \\
UA-Speech$_{5000}$      & 116.5    & 110.8     & 123.9          & 118.1 \\
CommonVoice$_{5000}$    & 16.5     & 15.9      & 45.5           & 35.2 \\
\midrule
\textbf{WER variance} \\
TORGO$_{5000}$          & 0.6      & 0.51       & 0.46          & 0.39  \\
UA-Speech$_{5000}$      & 0.58     & 0.50       & 0.76          & 0.67  \\
CommonVoice$_{5000}$    & 0.19     & 0.19       & 0.36          & 0.32  \\
\midrule
\textbf{PER, \%} \\
TORGO$_{5000}$          & 60.9     & 53.8       & 91.3          & 81.8 \\
UA-Speech$_{5000}$      & 91.0     & 88.3       & 103.5         & 94.5 \\
CommonVoice$_{5000}$    & 12.6     & 12.0       & 35.9          & 27.2 \\
\bottomrule
\end{tabular}
\newline
\newline
\textit{WER, WER variance and PER comparison for the ASR outputs on clean and transformed spectrograms across three speech dataset subsets}

\end{table}

\begin{table}[h]
\centering
\caption{WER and PER on problematic samples}
\label{tab:wer50}
\vspace{0.5em}
\begin{tabular}{l|cc|cc}
\toprule
\textbf{} & \multicolumn{2}{c|}{\textbf{RNNT-Conformer}} & \multicolumn{2}{c}{\textbf{CrisperWhisper}}  \\
& \textit{Identity} & \textit{Model v.2} & \textit{Identity} & \textit{Model v.2} \\   
\midrule
\textbf{WER, \%} \\
TORGO$_{>50\%}$          & 110.2    & 92.4     & 109.0     & 97.8 \\
UA-Speech$_{>50\%}$      & 123.6    & 114.5    & 125.9     & 119.4 \\
CommonVoice$_{>50\%}$    & 64.4     & 61.4     & 85.2      & 59.8 \\
\midrule
\textbf{PER, \%} \\
TORGO$_{>50\%}$          & 87.9     & 69.1      & 97.9      & 87.1 \\
UA-Speech$_{>50\%}$      & 97.1     & 91.3      & 105.6     & 93.2  \\
CommonVoice$_{>50\%}$    & 87.9     & 69.0      & 65.7      & 46.7 \\

\bottomrule
\end{tabular}
\newline
\newline
\textit{WER and PER comparison for the ASR outputs on clean and transformed spectrograms across high-difficulty subsets}
\end{table}

\subsection{Ablation Studies}

To better understand the contribution of individual components in our framework, we performed a series of ablation experiments, isolating specific transformation fields and evaluating their effects independently.

These ablation results support our core claim: the learned Lie-group fields provide ASR-aligned corrections that are:

\begin{itemize}
    \item Most effective when multiple transformation types are active,
    \item Non-harmful to clean speech and helpful even for non-pathological distortions,
    \item Distinct from random augmentation and statistically more effective,
    \item Structured, localized, and input-aware.
\end{itemize}

Detailed quantitative results, including WER and CER values for individual transformation modes and random warps, are provided in Appendix~\ref{app:ablation_details}.

\textbf{Do Learned Fields Perform Inversion?}
Although our model was originally motivated by the idea of learning approximate inverse transformations to dysarthric distortions, empirical results suggest that the predicted fields on \textit{clean} spectrograms are also non-zero and of similar norm to those predicted for dysarthric inputs. However, they do \textit{not degrade} ASR performance on normal speech, implying that these transformations, while active, do not introduce harmful artifacts. This suggests that the fields may not act as true inverses in a strict sense, but instead serve as \textbf{structurally beneficial corrections} aligned with ASR objectives.

\textbf{Per-Field Contribution.}
We evaluated the effect of each transformation field in isolation---time stretching, frequency scaling, 2D warping, and amplitude modulation---by running inference with only one active field at a time. Results show that:

\begin{itemize}
    \item Whole-dataset WER often remains statistically unchanged, with high p-values.
    \item Statistically significant gains are observed on a high-WER subset (samples where baseline WER $>$ 50\%).
\end{itemize}

This indicates that each field contributes positively in challenging cases, while full benefits emerge when multiple transformation types are combined.

We also observe that, depending on initialization and optimization dynamics, the model may converge to different \textit{dominant field configurations}. For example, the model trained with the symmetry-breaking potential (v2) consistently assigns highest energy to the amplitude field, suggesting that it captures the most informative structure for ASR normalization.

\textbf{Role of Random Transformations.}
To assess whether the observed improvements stem from learned structure or merely from distortion injection, we tested a baseline using \textit{randomly generated smooth Lie-style transformations}. These fields were drawn from the same distribution as the synthetic training distortions but applied directly at inference, without any prediction model. This setup resulted in measurable ASR gains, particularly on difficult dysarthric samples, confirming that structured local perturbations can improve robustness. However, performance remained consistently below the learned model, emphasizing the value of \textit{input-conditioned transformation fields}.

\textbf{Effect on Clean Speech and Generalization.}
Interestingly, we observed modest yet statistically significant improvements on the CommonVoice dataset, including high-WER samples. This suggests that the learned transformations may also help mitigate more general types of phoneme blur, temporal misalignments, or amplitude inconsistencies that occur in non-pathological speech and still challenge ASR systems. The large WER gains on CommonVoice using CrisperWhisper may partially result from the spectrogram upsampling step (80 to 128 mel bins) required to match Whisper's input format. This interpolation likely reduces quantization artifacts and unintentionally smooths the input, improving ASR robustness even on clean speech.

\section{Qualitative Analysis}

To better understand the behavior of the learned transformation fields and their impact on ASR, we conducted a qualitative analysis of predicted fields and their effects on spectrograms and transcriptions.

\subsection{Field Localization and Structure}

The predicted scalar fields exhibit clear spatial structure and energy-aware localization, shaped by the spectrogram content:

\begin{itemize}
    \item \textbf{Amplitude fields} show strong activation in regions with prominent spectral energy, often focusing around voice-active segments. They appear to modulate local intensity in ways correlated with energy contours.

    \item \textbf{Time-warping fields} are non-zero primarily in dynamic regions of the spectrogram, particularly where temporal structure varies. Their patterns suggest local adjustments in pacing or timing that are aligned with speech activity.

    \item \textbf{Frequency and 2D warp fields} display smooth, structured variation across the time--frequency grid. These patterns tend to reflect underlying spectral organization, such as banded or sloped energy distributions.

    \item \textbf{Low-energy and silent regions} are typically associated with minimal or smoothly varying field values, consistent with the sparsity constraints applied during training and the model’s tendency to avoid unnecessary deformation.
\end{itemize}

These qualitative behaviors suggest that the learned fields respond to input structure in meaningful ways, even in the absence of explicit supervision for phonemes or linguistic features. Visualizations are provided in Appendix~\ref{app:qualitative_figures}.

\subsection{Error Suppression and Fluency}

Manual inspection of ASR outputs before and after transformation reveals that the model frequently corrects characteristic errors on dysarthric speech. A notable pattern is the reduction of \textit{hallucinated word repetitions}, where baseline ASR predictions repeat a target word two or three times. These artifacts are often suppressed or eliminated in the transformed version.

Moreover, many reconstructed spectrograms exhibit visually cleaner phoneme boundaries and improved energy contrast—especially in consonant regions that tend to be weak in pathological speech.

\subsection{ASR Output Variability}

The transformations not only reduce word and character error rates but also lead to lower variance in ASR outputs across the dataset. This stabilization effect—combined with improved recognition of previously misclassified samples—suggests that the learned fields act as a kind of structured, interpretable front-end normalization.

\subsection{Appendix Visualizations}

We provide representative visualizations of predicted fields and their corresponding reconstructed spectrograms in Appendix~\ref{app:qualitative_figures}. These include:

\begin{itemize}
    \item Field maps overlaid on the input spectrogram,
    \item Original vs. reconstructed magnitude plots,
    \item ASR transcription comparisons for selected utterances.
\end{itemize}

These qualitative results reinforce the hypothesis that the learned Lie-group fields are meaningful, input-dependent, and linguistically aligned—even in the absence of explicit supervision.

\section{Limitations and Future Work}

While our framework demonstrates promising results in modeling and normalizing dysarthric speech via structured Lie group transformations, several limitations remain.

\textbf{Single-Mode Transformations.}  
In our current setup, only one transformation is applied per training sample. While this improves interpretability and code clarity, it restricts the model’s exposure to realistic compound distortions. Dysarthric speech often exhibits simultaneous time, frequency, and amplitude deviations. Future work could explore the application of multiple generators per sample—either jointly or sequentially—via compositional flows, mixture models, or recurrent multi-step training.

\textbf{Phase Ignorance.}  
We ignore phase information and operate purely on magnitude spectrograms. This limits the capacity to model voicing irregularities or pitch-related distortions that are primarily encoded in the STFT phase. Extending the model to act on complex-valued spectrograms or explicitly modeling phase derivatives (e.g., instantaneous frequency, group delay) could enable more precise and physically grounded corrections.

\textbf{No Explicit Phonetic or Linguistic Conditioning.}  
Our model is trained without knowledge of phonemes, articulatory features, or linguistic structure. While this ensures generality, it also limits targeted adaptation to distortion-prone phoneme classes or speaking styles. Future extensions could integrate weak supervision from phonetic aligners, forced alignment tools, or articulatory embeddings to better localize transformations where they matter most.

\textbf{Field Parameterization.}  
All transformation fields are predicted densely over the full time--frequency grid. In principle, many of these fields are low-rank or locally smooth. A sparse control grid or low-dimensional basis (e.g., radial basis functions, splines) could dramatically reduce model size, improve generalization, and enhance interpretability.

\textbf{Non-Uniqueness of Inversion.}  
Due to the infinite number of Lie field configurations that could lead to the same distorted spectrogram, the model's inverse is inherently non-unique. Rather than recovering the true deformation, it learns a consistent interpretation that aligns with observed structure in the data. Incorporating priors or task-specific constraints (e.g., articulatory models, speech intelligibility criteria) could steer learning toward more physiologically plausible or linguistically meaningful solutions.

\textbf{Loss Design.}  
Our training objective includes MSE, smoothness, cosine alignment, and sparsity terms. In the v2 model, we also introduce a spontaneous-symmetry-breaking (SSB) potential---formulated as a masked local “hat” loss---that encourages non-trivial field configurations and helps avoid convergence to zero-vacuum solutions. Future work could explore even more structured loss functions, including perceptually motivated metrics such as Mel Cepstral Distortion (MCD), or self-supervised ASR-based loss terms that better align with recognition accuracy.

\textbf{Spectrogram Inversion and Exponential Flow.}  
Our current inverse transformation is based on a first-order approximation of the exponential map. Extending this to higher-order expansions or iterative flows (e.g., via the Baker--Campbell--Hausdorff formula) could improve invertibility and enable modeling of more severe or non-local dysarthric distortions.

\textbf{Data Limitations and Paired Supervision.}  
The training pipeline currently relies on synthetic distortions applied to clean speech. Collecting or generating realistic paired datasets of \textit{(dysarthric, normalized)} spectrograms---via simulation, speech therapy corpora, or data augmentation pipelines---could provide stronger supervision and more effective field learning.

\textbf{Clinical Relevance and Personalization.}  
Our approach currently applies a generic normalization across speakers. In practice, dysarthria varies widely between individuals. Future extensions could incorporate speaker-specific adaptation, few-shot personalization, or use embeddings to condition the transformation fields on speaker identity.

We see these directions as natural extensions to our framework, all of which are supported by its Lie-theoretic foundations, differentiable operators, and modular loss structure.

\section{Conclusion}

We presented a geometry–based framework that treats dysarthric speech as the action of a local Lie group on spectrograms.  Time warping, frequency distortion, 2‑D warping, and amplitude modulation are cast as parameterised generator fields, yielding a single, differentiable front‑end that can be trained end‑to‑end from synthetic data and evaluated zero‑shot on real pathological speech.

Extensive experiments show that the learned, \emph{input‑conditioned} fields consistently improve ASR on severe TORGO and UA‑Speech utterances while leaving recognition of clean CommonVoice speech unchanged.  Beyond lowering mean WER and CER, the system also reduces performance \emph{variance} and curbs common failure modes such as word‑repetition hallucinations.  Ablations confirm that random but smooth Lie‑style warps provide only part of the benefit; content‑aware fields matter, with amplitude modulation emerging as the dominant corrective component in the v2 model. For reproducibility and further studies we make the open source code available at \url{https://github.com/miosipof/lie-dasr}. 

A spontaneous‑symmetry‑breaking (SSB) “hat’’ loss proved essential for avoiding collapse to the trivial zero‑field vacuum and enabled the model to learn non‑trivial, energy‑aligned deformation patterns at higher distortion strengths.  Field visualisations support the qualitative link between predicted transforms and spectro‑temporal energy structure.

\smallskip
\noindent\textbf{Future work.}  
The framework remains modular and naturally supports a range of extensions, including
(i) incorporating complex‑valued spectra to model phase,
(ii) employing higher‑order or iterative exponential flows for stronger inverses,
(iii) injecting weak phonetic or articulatory priors,
(iv) adopting sparse or low‑rank parameterisations of the fields,
(v) collecting or simulating realistic paired \emph{(dysarthric, clean)} corpora for supervised fine‑tuning, and
(vi) further exploring the geometric and physics‑inspired principles that underlie robust speech recognition.

\appendix

\textit{Note on Version 2.} 
This updated version of the paper expands substantially on the original submission. In addition to the original Lie group formulation and v1 model, we introduce an enhanced training strategy based on a spontaneous-symmetry-breaking (SSB) potential and masked local loss. A new v2 model is trained under this setup, yielding improved performance. We also conduct a series of ablation studies isolating individual transformation types, evaluate a randomized Lie-baseline, and report results using an additional ASR backend (CrisperWhisper) for robustness verification. Furthermore, we provide visualizations of the predicted transformation fields and spectrogram reconstructions, offering qualitative insights into the model’s behavior. All code and experiments are made available via open-source repository.

\section{Lie algebra}
\subsection{Algebra Properties}
\label{app:algebra_properties}

The set of generators used in our framework—local time/frequency warps, 2D deformations, and amplitude/phase modulations—defines an infinite-dimensional Lie algebra of differential operators with smooth scalar field coefficients. While these generators act locally and smoothly on spectrograms, their Lie brackets (commutators) do not, in general, close within the original generator set. For example, brackets between warping and amplitude operators yield higher-order differential terms (e.g., $\partial_t \alpha$, $\partial_f u_t$), indicating that the algebra extends beyond a finite basis.

This structure aligns with operator algebras studied in diffeomorphic registration, geometric mechanics, and continuous normalizing flows \cite{younes2019shape, marsden1999introduction, grathwohl2019ffjord}. While exact closure is not preserved, the exponential maps defining the transformations remain locally valid and differentiable. This is sufficient for training, inference, and geometric consistency within our framework.

The algebraic formulation is also compatible with extensions from geometric learning and field theory, where similar infinite-dimensional Lie algebras govern flows, gauge symmetries, and structured deformations \cite{bronstein2021geometric, kac1994infinite, cohen1995time}.

\section{Discretization}

\subsection{Sparse Field Parameterization and Control Grid}
\label{app:sparse_fields}

Although our current implementation uses dense transformation fields over the full \( F \times T \) spectrogram grid, one can quadratically reduce the number of parameters by defining the fields on a sparse grid and using interpolation.

Placing control points every $r$ steps in both time and frequency dimensions yields:
\[
\frac{F}{r} \times \frac{T}{r} = F \cdot T \times \frac{1}{r^2}  \text{ control points per field.}
\]

For example, in case of a $80 \times 512$ spectrogram, a reduction factor $r = 2$ would lead to $10240$ generator field values instead of a dense $80 \times 512 = 40960$ grid.

These sparse values can be interpolated to the full resolution using smooth basis functions such as 2D splines or Gaussian radial basis functions. Additionally, global structure can be modeled by adding low-degree polynomial terms in \( t \) and \( f \) to capture global slowdowns, frequency shifts, or articulatory drift.

Beyond parameter reduction, sparsity also aids in maintaining invertibility. Using monotonic splines or smoothly constrained fields helps prevent "folds" in the time or frequency axis that could arise from unconstrained dense fields. This improves numerical stability, makes inversion tractable, and ensures the transformations remain within the Lie group structure assumed by our model.

\subsection{Handling Large Distortions via Nonlinear Expansion}
\label{app:nonlinear_expansion}

In this work, we model spectrogram transformations using first-order Lie algebra actions:
\[
\widetilde S \approx S + \varepsilon X[S],
\]
which suffice for small to moderate distortions. However, for more severe or nonlinear cases, it is more accurate to use the full exponential map:
\[
\widetilde S = \exp(\varepsilon X)[S],
\]
or, in cases involving multiple non-commuting generators \( X_i \), the Baker--Campbell--Hausdorff (BCH) expansion:
\[
\widetilde S = \exp\left(\sum_i \gamma_i X_i + \frac{1}{2} \sum_{i,j} \eta_{ij} [X_i, X_j] + \dots \right)[S].
\]

This formulation allows large, structured transformations while preserving invertibility and the underlying Lie group geometry. Notably, it does not require any increase in the parameterization of the generator fields—the same fields \( \phi(f,t) \) may simply be applied for longer "flow time" \( \varepsilon \), or iteratively composed.

\section{Neural network}
\label{app:nnet}

\subsection{Model setup}
The model is configured as follows:
\begin{quote}
\texttt{backbone = smp.Unet(
encoder\_name="resnext50\_32x4d", 
encoder\_weights="imagenet", 
in\_channels=1, 
classes=5)}
\end{quote}

\subsection{Field Generation for Synthetic Training}
\label{app:field_generation}

To simulate realistic, smooth distortions in time and frequency, we use custom generators that create localized sinusoidal deformation fields. These fields are applied to magnitude spectrograms of normal speech to produce synthetic examples with known transformation parameters.

\paragraph{1D Fields.}  
Global time and frequency warp fields are constructed using sinusoidal functions masked by soft Gaussian windows:
\begin{itemize}
    \item A small number of sinusoidal “blobs” are placed along the 1D axis (time or frequency).
    \item Each blob has a randomly sampled frequency, phase, and amplitude.
    \item A soft mask confines the blob to a local region of the axis, enforcing smoothness and locality.
\end{itemize}

\paragraph{2D Fields.}  
Local time--frequency deformation fields use a similar principle in two dimensions:
\begin{itemize}
    \item We generate sinusoidal patterns over a 2D \([F, T]\) grid using random spatial frequencies and phases in both directions.
    \item Each wave is modulated by a soft 2D Gaussian mask centered at a random location.
    \item The final field is a sum of several such localized oscillatory components.
\end{itemize}

Both field types are parameterized by a small number of components (e.g., 2--3 blobs), allowing us to control smoothness and distortion strength via parameters like \texttt{mask\_radius\_frac} and \texttt{epsilon}. This procedure allows us to produce plausible distortions while retaining full control over ground-truth transformation parameters for supervision.

\subsection{Transformation Application and Curriculum Learning}
\label{app:transform_pipeline}

Given a clean spectrogram \( S(f,t) \), we apply one or more synthetic Lie algebra transformations to obtain a distorted spectrogram \( \widetilde{S}(f,t) \) used as input for training. Each transformation mode corresponds to a generator defined in our field construction pipeline.

\paragraph{Modes}
We support the following transformation types:
\begin{itemize}
    \item \texttt{t\_stretch}: local time warping (applied via warped time coordinate grid),
    \item \texttt{f\_stretch}: local frequency warping,
    \item \texttt{warp\_2d}: general 2D flow (independent vector fields for time and frequency),
    \item \texttt{amplitude}: smooth amplitude modulation,
    \item \texttt{phase} (optional): phase manipulation (currently not applied).
\end{itemize}

\paragraph{Sampling and Warping}
We use smooth sinusoidal fields with soft masks to generate the transformation fields. These are applied using bilinear warping based on additive coordinate shifts for each time--frequency bin:
\[
(f,t) \mapsto (f + \delta f, t + \delta t),
\]
where \( \delta f \) and \( \delta t \) are derived from the predicted or sampled scalar fields.

\paragraph{Curriculum Learning}
To progressively introduce more severe distortions, we define an \texttt{epsilon\_dict} specifying the magnitude of each generator. During early epochs, these values are small, producing near-identity transformations. As training progresses, \texttt{epsilon\_dict} is gradually increased, allowing stronger distortions. This strategy stabilizes learning while improving the model’s ability to generalize to severe dysarthric patterns.

\paragraph{Transform Composition}
While our system supports applying multiple transformations per sample (e.g., time warp + amplitude shift), we currently apply only one transformation per training example. This simplification facilitates interpretability and training stability. Future work may explore the benefits of transformation composition, which could more closely approximate real-world dysarthric variability.

\subsection{Inverse Transformation Pipeline}
\label{app:inverse_transform}

Given a distorted spectrogram \( S_{\text{distorted}} \) and a predicted set of normalized transformation fields \( \phi_{\text{pred}} \in \mathbb{R}^{B \times 5 \times F \times T} \), we compute an approximate inverse transformation to recover a normalized spectrogram \( S_{\text{recon}} \).

\paragraph{Rescaling Fields}
Each predicted field is denormalized using the predefined maximum distortion levels \(\varepsilon_{\text{type}}\) as specified in a dictionary:
\[
\phi_{\text{real}} = \phi_{\text{pred}} \cdot \varepsilon_{\text{type}}
\]

\paragraph{Inversion Steps}
We apply the inverse of each transformation sequentially:
\begin{enumerate}
    \item Reverse global and local time/frequency warps using negated displacement fields.
    \item Reverse 2D warps using the same method with independent flows for time and frequency.
    \item Undo amplitude modulation using the inverse operator \( S / (1 + \alpha) \), clamped for safety to avoid division by zero.
\end{enumerate}

This pipeline is implemented efficiently using PyTorch and preserves differentiability if needed for future applications in backpropagation-based learning. While this is an approximation of the true group inverse, it is sufficient for training purposes and helps to enforce consistency between the predicted and ground-truth transformation space.

\subsection{Model v.1 Training}
\label{app:training_v1}

The first version of the model was trained for 10 epochs on a 10,000-sample subset of the English CommonVoice dataset using distributed data parallelism (DDP) across four NVIDIA GeForce RTX 4090 GPUs. We used an initial learning rate of $3 \times 10^{-5}$ with exponential decay (decay factor $\gamma = 0.9$), a batch size of 32, and an Adam optimizer.

Training followed a simple curriculum schedule governed by the distortion strength parameter $\varepsilon$, which linearly increased from 0.5 to 4.0 over the training, with a plateau at $\varepsilon = 2.0$ maintained during epochs 4--6. The loss function combined mean squared error (MSE) between predicted and ground-truth fields, cosine similarity, and spectrogram reconstruction loss.

\begin{figure}[ht!]
\centering
\includegraphics[width=80mm]{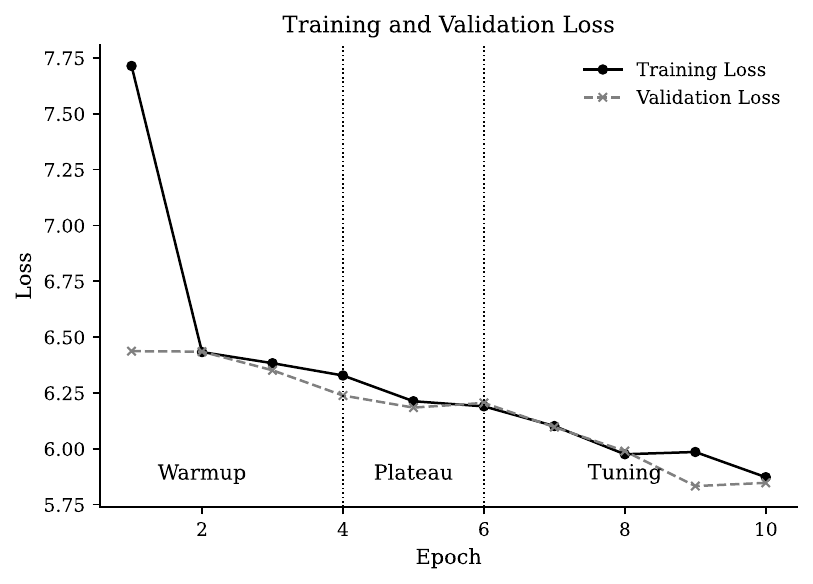}
\caption{Training and validation loss dynamics (10000 samples from CommonVoice dataset. Model version v.1, batch size = 32, learning rate starts from $3 e^{-5}$). The $\varepsilon$ parameter grows linearly on warmup stage, plateaus and then grows linearly until the end of training}
\label{fig:train_loss_v1}
\end{figure}

\begin{figure}[ht!]
\centering
\includegraphics[width=80mm]{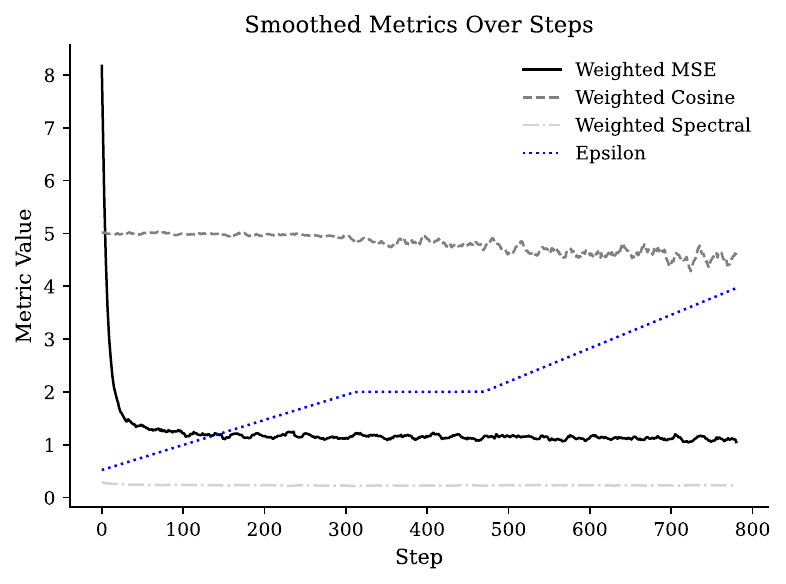}
\caption{Weighted loss function terms and $\varepsilon$ dynamics across training steps. Model version v.1}
\label{fig:loss_components_v1}
\end{figure}

Figure~\ref{fig:train_loss_v1} shows the training and validation loss curves over time. Figure~\ref{fig:loss_components_v1} presents the evolution of individual weighted loss components (field MSE, cosine similarity, and spectrogram reconstruction MSE) along with the $\varepsilon$ schedule across training steps.

\subsection{Model v.1 Evaluation on TORGO}
\label{app:results_v1}

We evaluated the v1 model on a held-out subset of 5,000 randomly selected samples from the TORGO dataset labeled as ``dysarthric.'' Evaluation was conducted using the Conformer-Transducer ASR backend. Table~\ref{tab:wer_v1} summarizes word error rate (WER) and character error rate (CER) for clean and transformed spectrograms across the full evaluation set. Table~\ref{tab:wer_v1_50} further reports performance on subsets of ``high-difficulty'' samples—defined as those where the untransformed spectrogram yields WER~$>$~50\% or CER~$>$~30\%. The v1 model shows consistent improvement in both WER and CER, with particularly notable gains on the high-WER subset, suggesting its capacity to selectively normalize difficult speech segments without degrading performance on cleaner examples.

\begin{table}[h]
\centering
\caption{WER and CER across different datasets and spectrogram transformations, model v.1}
\label{tab:wer_v1}
\vspace{0.5em}
\begin{tabular}{l|cc|cc|cc}
\toprule
\textbf{} & \multicolumn{2}{c|}{\textbf{TORGO}} & \multicolumn{2}{c|}{\textbf{UASpeech}} & \multicolumn{2}{c}{\textbf{CommonVoice}} \\
                   & \textit{WER} & \textit{CER}        & \textit{WER} & \textit{CER}           & \textit{WER} & \textit{CER} \\
\midrule
Clean    &  77.2 & 63.1  &  116.5 & 88.3  &  16.4 & 12.3  \\
v.1  &  73.9 & 61.1  &  112.4 & 89.6  &  16.2 & 11.9  \\
\bottomrule
\end{tabular}
\newline
\newline
\textit{WER and CER comparison for the ASR outputs on clean and transformed with model v.1 spectrograms across three speech dataset subsets (5{,}000 each, "medium" severity selected from UASpeech dataset)}
\end{table}

\begin{table}[h]
\centering
\caption{WER and CER on problematic samples, model v.1}
\label{tab:wer_v1_50}
\vspace{0.5em}
\begin{tabular}{l|cc|cc|cc}
\toprule
\textbf{} & \multicolumn{2}{c|}{\textbf{TORGO$_{>50}$}} & \multicolumn{2}{c|}{\textbf{UASpeech$_{>50}$}} & \multicolumn{2}{c}{\textbf{CommonVoice$_{>50}$}} \\
                   & \textit{WER} & \textit{CER}         & \textit{WER} & \textit{CER}           & \textit{WER} & \textit{CER} \\
\midrule
Clean    &  110.1 & 91.1  &  123.6 & 96.4  & 64.4 & 41.7  \\
v.1      &  93.7  & 76.8  &  115.7 & 94.0  & 62.3 & 40.6  \\
\bottomrule
\end{tabular}
\newline
\newline
\textit{WER and CER comparison for the RNNT Conformer ASR outputs on clean and transformed with model v.1 spectrograms across high-difficulty subsets}
\end{table}

\subsection{Ablation Studies - Details}
\label{app:ablation_details}

\textbf{Ablation: Single-Generator Models.}
To assess the contribution of each transformation type, we trained four separate ablation models—each with only one active field generator (time stretch, frequency warp, 2D warp, or amplitude modulation)—following the same training protocol as the v1 model (10 epochs on 10{,}000 CommonVoice samples, exponential LR decay). When evaluated on the TORGO dataset, none of the single-generator models yielded statistically significant WER improvements on the full 5{,}000-sample test set. However, moderate but consistent gains were observed on high-WER and high-CER subsets (defined as WER $>$ 50\% or CER $>$ 30\%), with amplitude-only transformations being the most effective. These results are reported in Table~\ref{tab:single_fields}. Despite these partial gains, all single-generator models underperformed relative to the full v1 model trained with access to all transformation types, suggesting that the interplay of multiple Lie-group generators is critical for capturing the complexity of dysarthric distortions.

\textbf{Ablation: Random Smooth Transformations.}

To evaluate the contribution of learned, input-dependent transformation fields, we conducted a baseline experiment in which smooth but random Lie-style fields were applied instead of using predicted $\phi$ fields. These fields were sampled from the same sinusoidal generator used during training but without conditioning on the input spectrogram. Four separate evaluations were performed, each using only one type of transformation (time, frequency, 2D warp, or amplitude) applied independently to dysarthric speech from the TORGO dataset. Across the full test set, none of the transformations led to statistically significant WER changes across the whole dataset (small gains are associated with high $p$-values). However, moderate improvements were consistently observed on the high-WER and high-CER subsets, indicating that the geometric nature of the transformations alone can be beneficial in some cases. These results are reported in Table~\ref{tab:random_lie} of the Appendix. Nevertheless, the learned fields clearly outperform random ones, highlighting the value of input-conditioned warping.

\subsection{ASR Decoding Parameters for CrisperWhisper}

For experiments using the CrisperWhisper ASR backend, we applied deterministic decoding with constrained sampling to minimize hallucinations and ensure consistency across evaluations. The following decoding parameters were used:

\begin{itemize}
    \item \texttt{do\_sample=False}: disables stochastic sampling.
    \item \texttt{num\_beams=1}: equivalent to greedy decoding.
    \item \texttt{temperature=0.0}: enforces deterministic token selection.
    \item \texttt{top\_k=1}, \texttt{top\_p=1.0}: disables top-$k$ and nucleus sampling.
    \item \texttt{repetition\_penalty=1.2}: discourages token repetition in the output.
    \item \texttt{length\_penalty=1.0}: applies no explicit bias toward shorter or longer sequences.
    \item \texttt{forced\_decoder\_ids}: ensures the correct language and task tokens are used.
\end{itemize}

These parameters were chosen to reflect CrisperWhisper's verbatim-mode training objective and to ensure alignment with the short, clearly segmented utterances found in dysarthric speech datasets.

\subsection{Qualitative Analysis - Figures}
\label{app:qualitative_figures}

\begin{figure}[ht!]
\centering
\includegraphics[width=80mm]{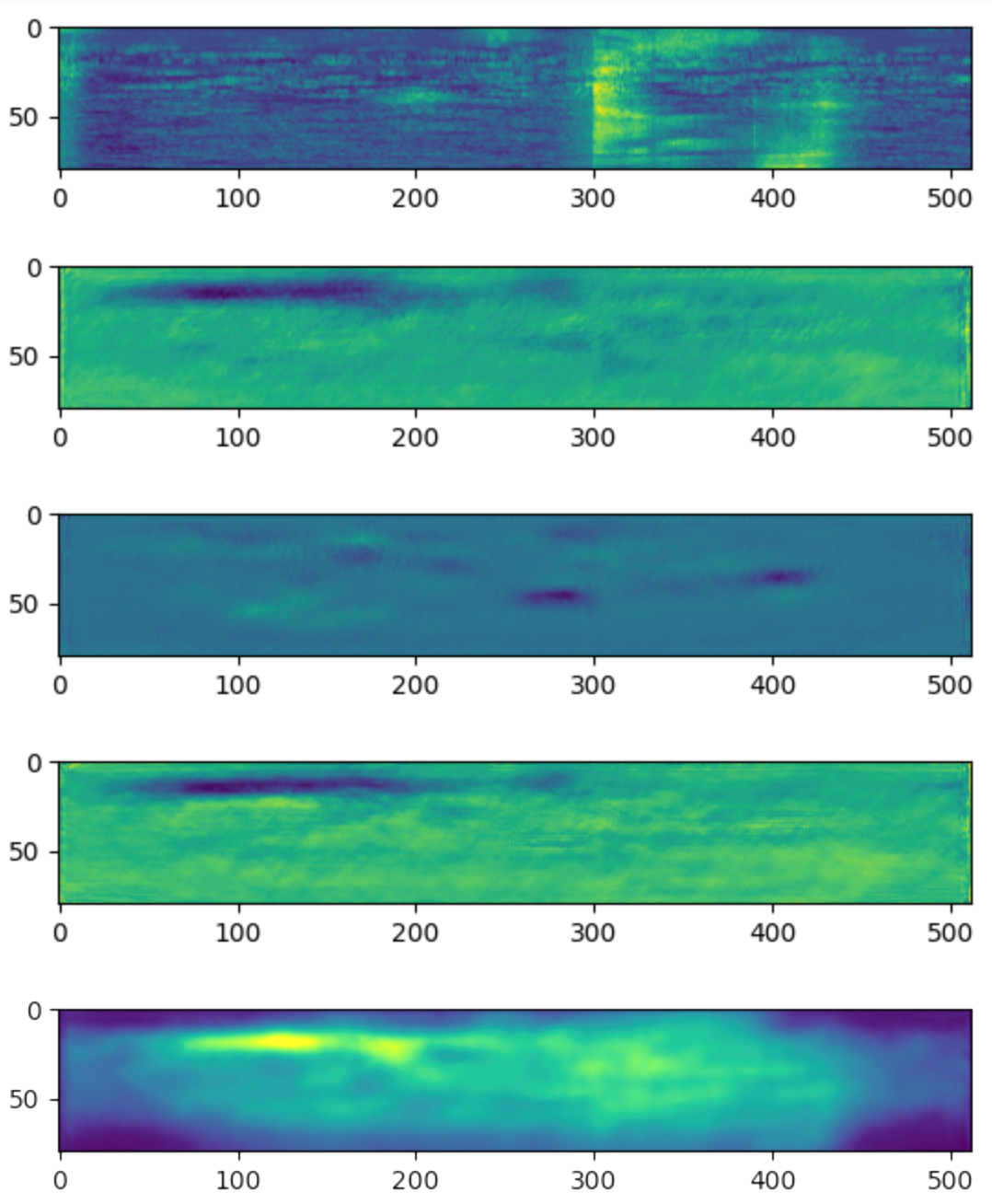}
\caption{Example of input spectrogram (first picture from above) and predicted fields: $\phi_{\text{time}}$, $\phi_{\text{freq}}$, $\phi_{\text{warp}_t}$, $\phi_{\text{amp}}$.
Original transcription: "peace". Transcription with transformed spectrogram: "pace". True label: "paste"
}
\label{fig:fields_pred}
\end{figure}

\begin{table}[h]
\centering
\caption{WER across different transformation generators}
\label{tab:single_fields}
\vspace{0.5em}
\begin{tabular}{l|c|c|c|c}
\toprule
\textbf{} & \textbf{TORGO} & \textbf{UA-Speech} & \textbf{TORGO$_{>50}$} & \textbf{UA-Speech$_{>50}$} \\
\midrule
Identity            &  77.3 &  116.5 &  110.1 &  123.6 \\
\midrule
$v\partial_t$       &  77.1 &  115.6 &  109.3 &  121.9 \\
$w\partial_f$       &  77.7 &  116.3 &  108.7 &  122.8 \\
Warp 2D             &  77.3 &  116.1 &  108.7 &  122.7 \\
$\alpha$            &  77.0 &  115.9 &  104.6 &  121.2 \\
\midrule
Model v.1          &  73.9  &  112.4 &  93.7  &  115.7 \\
\bottomrule
\end{tabular}
\newline
\newline
\textit{WER comparison for the RNNT Conformer ASR outputs on transformed spectrograms obtained with model versions (v.1) limited to a selected transformation type. Each model version trained for 10 epochs}
\end{table}

\begin{table}[h]
\centering
\caption{WER across different random smooth transformations}
\label{tab:random_lie}
\vspace{0.5em}
\begin{tabular}{l|cc|cc}
\toprule
\textbf{} & \multicolumn{2}{c|}{\textbf{TORGO}} & \multicolumn{2}{c}{\textbf{TORGO$_{>50}$}} \\

& \textit{WER} & \textit{CER} & \textit{WER} & \textit{CER} \\

\midrule
Identity            &  77.3 &  63.1 &  110.2 &  91.0 \\
\midrule
$v\partial_t$       &  77.1 &  63.0 &  109.1 &  90.0 \\
$w\partial_f$       &  77.7 &  62.9 &  107.0 &  87.9 \\
Warp 2D             &  77.3 &  63.0 &  108.6 &  89.6 \\
$\alpha$            &  77.0 &  62.4 &  106.7 &  87.4 \\
\midrule
Model v.2           &  72.3 &  57.6 &  92.4  &  73.7 \\
\bottomrule
\end{tabular}
\newline
\newline
\textit{WER comparison for the RNNT Conformer ASR outputs on transformed spectrograms obtained with random Lie-type transformations with $\varepsilon = 0.6$ acting on spectrograms from TORGO dataset and outputs on the spectrograms transformed with model v.2 with $\varepsilon = 0.6$}
\end{table}

\section{Open-Source Code}

To support reproducibility and further research, we release the full source code used for training, inference, and evaluation of our Lie transformation framework. The repository includes model definitions, data preprocessing scripts, training configurations, and utilities for ASR-based evaluation with multiple backends. The code is available at: \url{https://github.com/miosipof/lie-dasr}. Please note that this is a research-oriented implementation intended for academic use, and may not be optimized for production deployment.



\end{document}